\documentclass[sigconf]{acmart}

\usepackage{booktabs}
\usepackage{graphicx}
\usepackage{balance}
\usepackage{xspace}
\usepackage{caption}
\usepackage[T1]{fontenc}
\usepackage{colortbl}
\usepackage{listings}
\usepackage{xcolor}
\usepackage{enumerate}
\usepackage{framed}
\newcommand{\eg}{e.g.,\xspace}
\newcommand{\etc}{etc.\xspace}

\newcommand{\ie}{i.e.,\xspace}

\newcommand{\adstxt}{\texttt{ads.txt}\xspace}
\newcommand{\sellersjson}{\texttt{sellers.json}\xspace}

\newcommand{\cop}{CoP\xspace}

\newcommand{\point}[1]{\par\smallskip\noindent\textbf{#1}}

\copyrightyear{2025}
\acmYear{2025}
\setcopyright{cc}
\setcctype{by}
\acmConference[WWW '25]{Proceedings of the ACM Web Conference 2025}{April 28-May 2, 2025}{Sydney, NSW, Australia}
\acmBooktitle{Proceedings of the ACM Web Conference 2025 (WWW '25), April 28-May 2, 2025, Sydney, NSW, Australia}
\acmDOI{10.1145/3696410.3714898}
\acmISBN{979-8-4007-1274-6/25/04}

\begin{document}

\title{Before \& After: The Effect of EU's 2022 Code of Practice on Disinformation}

\author{Emmanouil Papadogiannakis}
\affiliation{
	\institution{\normalsize FORTH \& University of Crete}
        \city{Heraklion}
	\country{Greece}
}
\author{Panagiotis Papadopoulos}
\affiliation{
	\institution{\normalsize FORTH}
        \city{Heraklion}
	\country{Greece}
}
\author{Nicolas Kourtellis}
\affiliation{
	\institution{\normalsize Telefonica Research}
        \city{Barcelona}
	\country{Spain}
}
\author{Evangelos Markatos}
\affiliation{
	\institution{\normalsize FORTH \& University of Crete}
        \city{Heraklion}
	\country{Greece}
}

\renewcommand{\shortauthors}{Emmanouil Papadogiannakis, Panagiotis Papadopoulos, Nicolas Kourtellis, Evangelos Markatos}



\begin{abstract}

Over the past few years, the European Commission has made significant steps to reduce disinformation in cyberspace.
One of those steps has been the introduction of the 2022 ``Strengthened Code of Practice on Disinformation''.
Signed by leading online platforms, this Strengthened Code of Practice on Disinformation is an attempt to combat disinformation on the Web.
The Code of Practice includes a variety of measures including the demonetization of disinformation, urging, for example, advertisers  ``to avoid the placement of advertising next to Disinformation content''.

In this work, we set out to explore what was the impact of the Code of Practice and especially to explore to what extent ad networks continue to advertise on dis-/mis-information sites.
We perform a historical analysis and find that, although at a hasty glance things may seem to be improving, there is really \emph{no significant reduction in the amount of advertising relationships among popular misinformation websites and major ad networks}.
In fact, we show that ad networks have withdrawn mostly from \emph{unpopular} misinformation websites with very few visitors, but still form relationships with highly unreliable websites that account for the majority of misinformation traffic.
To make matters worse, we show that ad networks continue to place advertisements of legitimate companies next to misinformation content.
We show that major ad networks place ads in almost 400 misinformation websites in our dataset.

\end{abstract}

\maketitle

\section{Introduction}

From the US presidential elections in 2016~\cite{bovet2019influence} to the Brexit referendum~\cite{marshall2018post}, Covid-19~\cite{rocha2021impact}, the war in Ukraine~\cite{khaldarova2020fake} and the recent conflict in the Middle East~\cite{middleeast}, disinformation today is shaping major global events. It tears apart the fabric that holds our societies together by destroying people’s faith in traditional news sources and undermining people’s trust in governments, public institutions, and democratic processes~\cite{ognyanova2020misinformation}. Disinformation is usually designed to appeal to our worst impulses, fears and prejudices – all in an attempt to shape opinions or divide and conquer a society on the information battlefield. 
Over 80\% of EU citizens see fake news as an issue both for their country, and for democracy in general~\cite{fakenews_survey2018}.

In an attempt to mitigate this threat, governments, tech firms, and stakeholders have explored various methods to identify and curtail the spread of fake news.
In 2018, Google launched the Digital News Innovation Fund that supports over 7000 news partners (\$300M in funding)~\cite{dnifund}.
In 2022, the United Nations in their report~\cite{un_report} set out the relevant international legal framework, and discussed measures that States and technology enterprises reported to have taken to counter disinformation.
A few years earlier, the European Union issued the 2018 Code of Practice on Disinformation, where representatives of online platforms, leading tech companies and players in the advertising industry (\eg Facebook, Twitter, Google with Microsoft and TikTok following) agreed on a self-regulatory Code of Practice to address the spread of online disinformation~\cite{cop2018}.
The Code of Practice is based on 21 commitments in different domains, including the demonetization of purveyors of disinformation.

A year after its implementation, the Code was already acknowledged globally as a pioneering framework~\cite{covidAssessment}.
An assessment of the Code was published in 2020~\cite{assessment}, showing that (i) the Code did provide a valuable framework for a structured dialogue between online platforms, thus, ensuring transparency and accountability of their policies on disinformation, but also (ii) a set of important gaps and shortcomings.
In June 2022, the Commission issued the Strengthened Code of Practice on Disinformation (\cop)~\cite{codeOfPractice}.

In this paper, we set out to explore the impact of the Strengthened \cop with regard to one of its most important commitments: the demonetization of disinformation-spreading websites. We perform a historical analysis on misinformation websites in order to investigate what was the impact of the policy on their ad revenue by comparing their direct business relationships with ad networks.

The contributions of this work are summarized as follows:
\begin{enumerate}
    \item We discover that there is no significant reduction in the amount of advertising accounts with popular ad networks, after the Code of Practice on Disinformation came into effect. Popular ad networks have withdrawn mostly from unpopular disinformation websites, with hardly any visitors. On the contrary, we find that these ad networks still have connections with popular misinformation websites that account for the majority of misinformation traffic.
    \item Two years after the \cop was signed, signatories still accept 1 out of 3 misinformation websites as authorized ad inventory sellers.
    \item We show that ad networks do not substantially differentiate between unreliable and reliable news websites in their business relations, and that they form business relationships with both kinds of websites.
    \item We establish that ad networks still serve ads to almost 400 misinformation websites (from our dataset) and facilitate the monetization of misinformation content.
    \item We find that ads of 23 Fortune 500 companies are served next to misinformation content, endangering their brand reputation and consumer trust.
    \item We make our list of approximately 2,500 misinformation news websites publicly available~\cite{openSourceData}.
\end{enumerate}

\section{Background}
\label{ref:background}

When users visit a website, an automated programmatic process matches them with advertisers.
Advertisers rely on Demand-Side Platforms (DSPs) to target the right users, while website publishers use Supply-Side Platforms (SSPs) to manage their ad inventory.

\begin{figure*}[t]
    \begin{minipage}[t]{0.32\textwidth}
        \centering
        \includegraphics[width=\columnwidth]{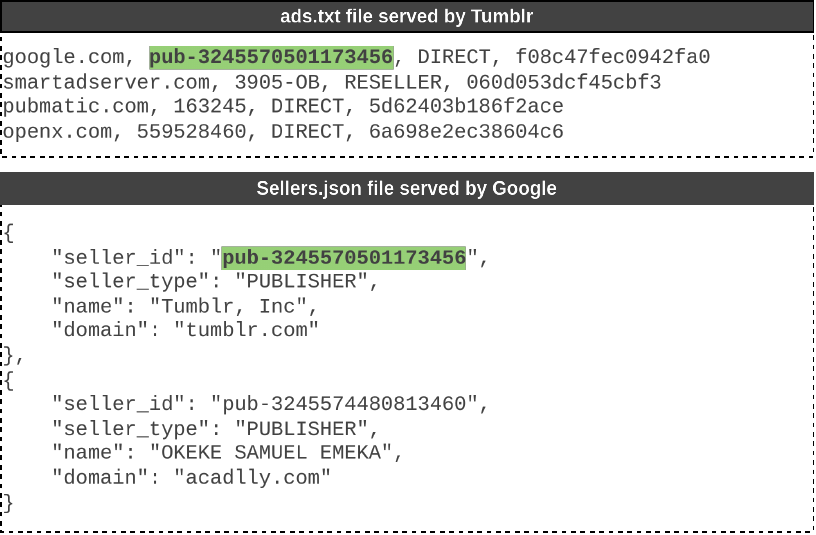}
        \caption{Snippets of Tumblr's \adstxt file and Google's \sellersjson file on April 2024. The ID that describes an advertising account matches in both files.}
        \label{fig:adstxt_sellersjson}
    \end{minipage}
    \hfill
    \begin{minipage}[t]{0.32\textwidth}
        \centering
	\includegraphics[width=\columnwidth]{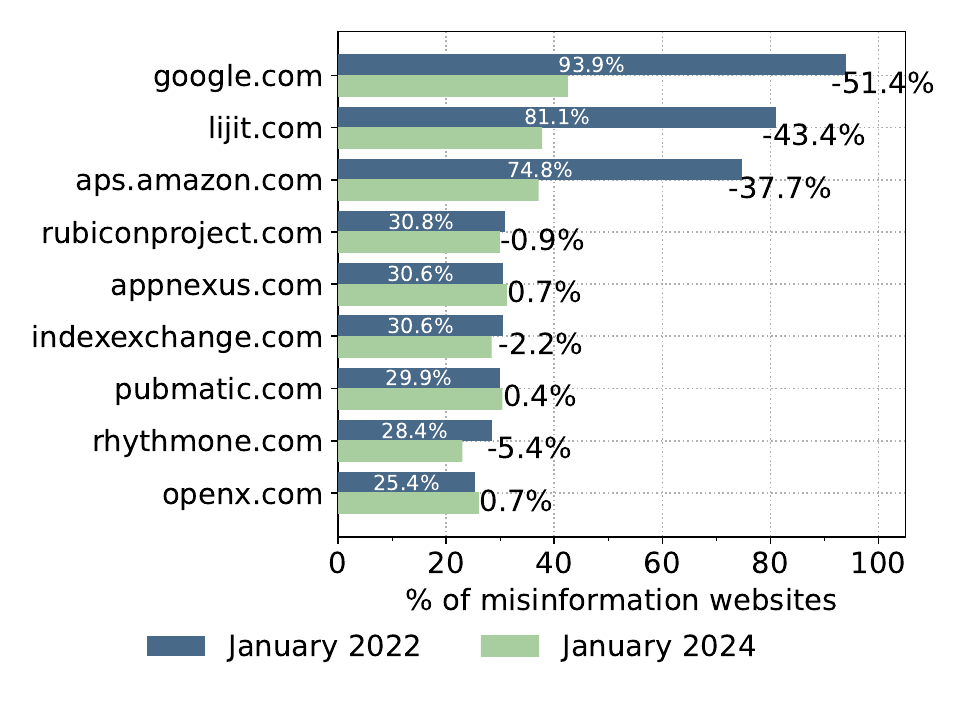}
	\caption{Direct business relationships of top ad networks with misinformation websites over the years.}
	\label{fig:HistoricAdsTxtBusinessRelationships}
    \end{minipage}
    \hfill
    \begin{minipage}[t]{0.32\textwidth}
        \centering
    	\includegraphics[width=\columnwidth]{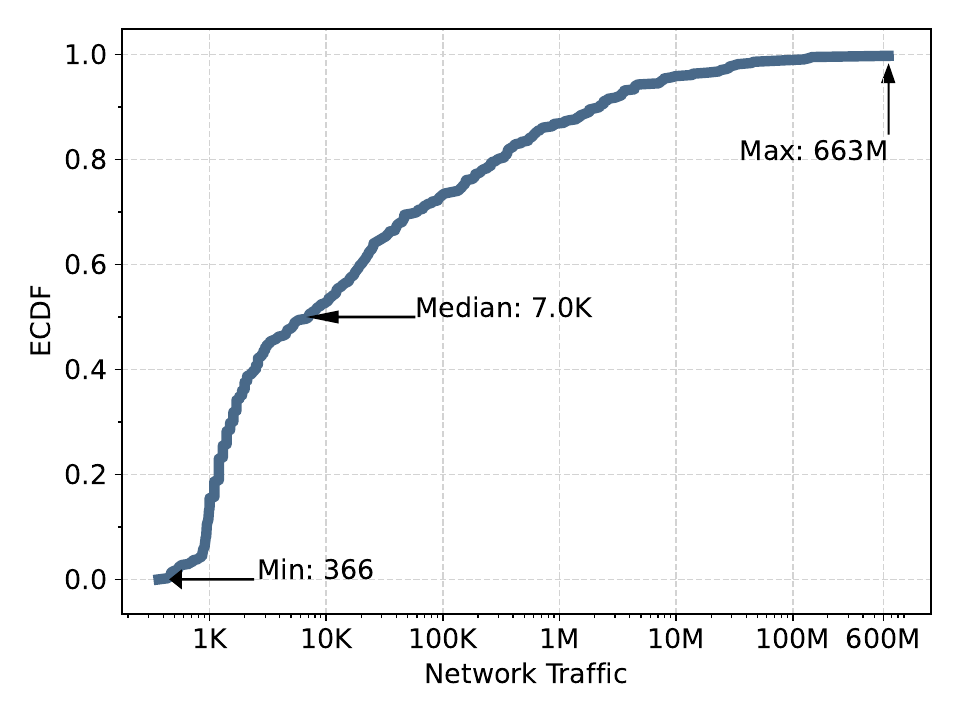}
    	\caption{Distribution of number of visits of selected misinformation websites on January 2024 (Note: $x$-axis is in log scale).}
    \label{fig:MisinformationWebsitesNetworkTraffic}
    \end{minipage}
\end{figure*}

\point{\adstxt:~}
Due to the complexity of the advertising ecosystem, bad actors were able to sell ad inventory of websites they did not control.
To eliminate this, the Internet Advertising Bureau (IAB) Technology Laboratory introduced the \adstxt specification~\cite{adsTxtStandard} that allows publishers to explicitly disclose who is authorized to sell the ad inventory of their website.
This information is enclosed within a text file, publicly available at the root of the domain (snippet in Figure~\ref{fig:adstxt_sellersjson}).
Each \adstxt file contains comma-separated records, with each record authorizing an entity to sell the ad inventory of the website.
Each record contains:
(i) the domain name of the advertising system (\eg SSP),
(ii) an identifier for the seller's account,
(iii) the type of the account, and optionally
(iv) a certification ID for the advertising system.
The type of the account represents the relationship between the account holder and the ad system.
A \texttt{DIRECT} account suggests that the website publisher also controls the advertising account and has a direct business contract with the ad system.
For a \texttt{RESELLER} account, the website publisher has authorized a third-party entity to resell their ad inventory.

\point{\sellersjson:~}
To increase transparency in the advertising ecosystem, the IAB Tech Lab also introduced the \sellersjson standard~\cite{sellersJsonStandard}.
This standard helps discover the identities of the entities that sell the ad inventory of websites and can help identify all intermediaries involved in ad inventory selling.
The \sellersjson standard is implemented as a JSON file published by advertising systems in order to disclose the inventory sellers they have approved within their ad system.
Each approved seller is represented by a unique identifier inside this file.
This is the same identifier that appears in the \adstxt file and maps to a single entity that is paid for the ad inventory it sells (example in Figure~\ref{fig:adstxt_sellersjson}).
Additionally, a seller type describes each identifier.
A \texttt{PUBLISHER} seller suggests that this account directly owns and controls the website whose ad inventory is sold through the advertising system, and that the advertising system directly pays this entity.
An \texttt{INTERMEDIARY} record indicates that it is a reseller of ad inventory.
Finally, a seller can be labelled as \texttt{BOTH}, acting as both a publisher and an intermediary.

\section{Methodology}

\subsection{Sources of Misinformation}
\label{sec:misinformation-websites}

Disinformation is the deliberate creation or spread of false information, whereas misinformation is false information shared without necessarily intending to deceive.
While we study the Code of Practice on Disinformation, our focus is on how misinformation websites generate ad revenue.
Proving that there is clear intent to mislead (\ie disinformation) is not a trivial task and falls outside the scope of this work.
Nonetheless, some websites in our list have been labeled as disinformation spreaders (\eg~\cite{mbfcWarRoom,mbfcTrendingPolitics}).
We acknowledge the distinction between these two concepts but argue that understanding the business model behind false content is crucial, as many disinformation campaigns rely on the same advertising mechanisms as misinformation~\cite{googleDisinfo,newsGuardDisinfo}.
The ability to profit from misleading content encourages bad actors to deliberately spread false claims, blurring the line between misinformation and disinformation.

To assemble a list of misinformation websites and study their behavior over time, we make use of publicly available datasets produced by academic publications, as well as by credible and acclaimed organizations.
Our first source is MediaBias/FactCheck (MBFC)~\cite{mbfc}, an organization that investigates bias and misinformation in online sources of information.
By using human evaluators and a concrete and consistent methodology~\cite{mbfcMethodology}, they evaluate the 
factual accuracy of various media sources.
We visit the MBFC website in October 2023 and extract labels and information about the bias, factual reporting, credibility and traffic volume for each evaluated website. We retain only websites that have a ``LOW CREDIBILITY'' score \textbf{and} a factual reporting of ``MIXED'', ``LOW'' or ``VERY LOW'', forming a list of 1,677 misinformation websites.
We consciously exclude websites from the ``Satire'' category from this analysis, since they deliberately use humor and exaggeration in their published articles and do not attempt to deceive visitors.
Such websites do not fall under the misinformation umbrella.

Next, we use the publicly available lists of previous academic works.
We extend our dataset with fake news websites assembled in~\cite{papadogiannakis2023funds} consisting of 1,044 misinformation websites.
Additionally, we extract 669 more misinformation sites from~\cite{vekaria2023inventory} that classifies news websites using independent sources (\eg Politifact, Snopes).
Finally, we make use of the \textit{FakeNewsNet} tool used in~\cite{shu2020fakenewsnet} and download all news articles from the collected dataset.
We exclude all data from the GossipCop fact-checking service because it has been discontinued and is no longer available.
We extract 8 misinformation websites by filtering websites that have published at least 4 fake news articles (proven by Politifact~\cite{politifact}).
We also filter out Facebook because we argue that it is neither a news publisher nor a news aggregator. 
Misinformation posts on a large platform shouldn't label the entire network as misinformation.

We merge all the above lists and end up with a dataset of 2,469 distinct misinformation websites.
In an attempt to foster academic research and give back to the academic community, 
we make our list of misinformation websites publicly available~\cite{openSourceData}.

\subsection{Fetching \adstxt and \sellersjson files}
\label{sec:dataset}

Our study revolves around \adstxt and \sellersjson, two important files in digital advertising that help combat ad fraud and increase transparency in the advertising ecosystem.
Previous work has demonstrated that these files can reveal business relationships among websites and ad networks~\cite{bashir2019longitudinal,papadogiannakis2023funds,vekaria2023inventory}.
We first study the business relationships among ad networks and misinformation websites and how these have changed over the years and especially after the Code of Practice on Disinformation was signed.
To that extent, we utilize \adstxt files served by misinformation websites, and the \sellersjson files served by popular ad networks.
We crawl the list of approximately 2,500 misinformation websites on January 2024 and download 226,659 \adstxt entries coming from 1,132 misinformation websites (an \adstxt file of a single website can contain multiple entries for multiple ad networks~\cite{papadogiannakis2023pooling}).
We make use of an open-source crawler~\cite{papadogiannakis2023pooling} to collect and parse the respective \adstxt files.
We find there are \adstxt records representing relationships with over 1,100 distinct ad networks, suggesting that ad networks still work with objectionable websites.
To get a better understanding of the actual business relationships, we direct our analysis to \adstxt entries marked as \texttt{DIRECT}.
The \adstxt specification defines that \texttt{DIRECT} accounts indicate the website owner (\ie publisher) directly controls the advertising account, and that there is a direct business contract between the publisher and the ad system~\cite{adsTxtStandard}.
Note that a website may provide bogus information in its \adstxt file.
For example, it may claim that a legitimate advertiser sells ads on this website.
On the other hand, an ad seller may also provide a fake \sellersjson file claiming that it can put ads in high-profile websites to attract more customers.
Although each individual file of the above two files may provide fake information, their intersection provides the truth.
Indeed, if website A (in its \adstxt file) claims that it displays content from ad network B, \textbf{and} ad network B (in its \sellersjson file) claims that it displays ads in website A, then this is true.
We provide an example in Figure~\ref{fig:adstxt_sellersjson}.
In the rest of this work we take the intersection of these files. 
That is, we form a relation between A and B only if we find this relation both in \adstxt file of A and in \sellersjson file of B. 

\subsection{Ad Collection}
\label{sec:ad-collection}

In order to study ads that are actually served on misinformation websites, as well as the ad networks that serve them, we develop a fully automated system that visits websites, emulates the behavior of a real user and clicks on ads.

First, we establish different user profiles (\ie personas) to emulate real users and present a compelling profile to ad networks.
We form different personas emulating different real-world users so that they receive ads from different advertisers, thus revealing a better picture of the ad ecosystem around misinformation.
Additionally, we emulate users from various geographic locations in order to (potentially) get ads from both acclaimed companies and brands known worldwide, as well as local advertisers.
We select Netherlands, Greece, and USA as the different vantage points.
We choose two EU countries since the \cop is an EU Commission initiative. Given its self-regulatory nature and potential differences outside the EU, we also include the US to identify ad servings that may not be evident in the EU.
We use a VPN service to set the location of our virtual users, and before we begin forming the personas, we use two different geolocation services to ensure that these virtual users indeed appear to be in the specified country.
In order to build a persona, we visit a set of specific websites and expect that third parties will infer the user’s gender and preferences based on the content they consume.
We describe the methodology we follow to build user profiles in Appendix~\ref{sec:personas}.

\begin{figure*}[ht]
	\begin{minipage}[t]{0.32\textwidth}
    	\centering
    	\includegraphics[width=1\columnwidth]{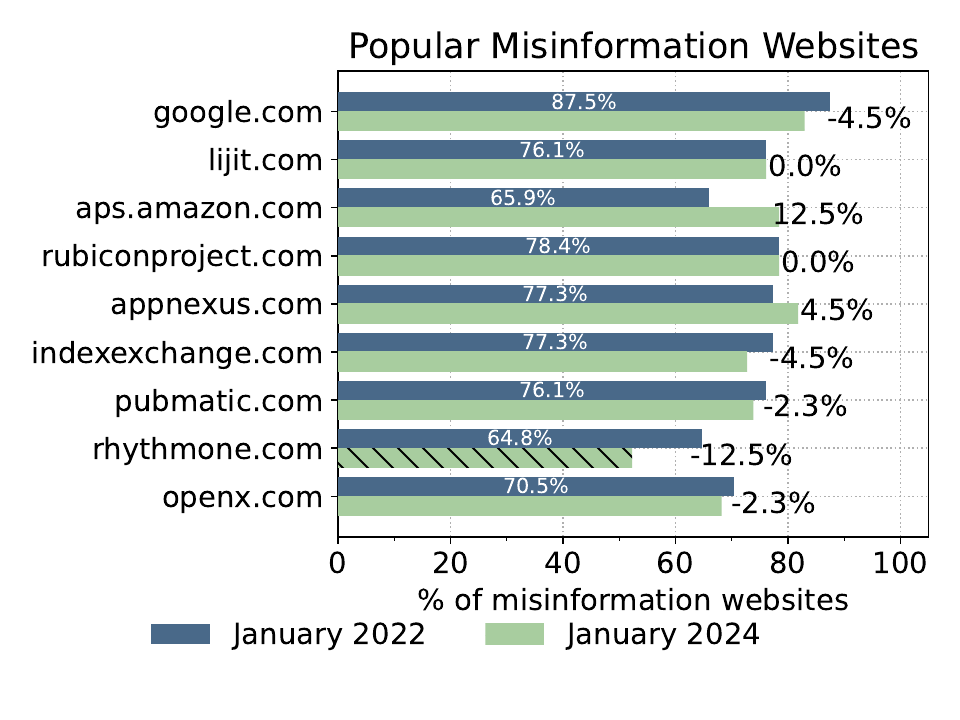}
    	\caption{Direct business relationships of top ad networks with \textbf{popular} misinformation websites over the span of 2 years.}
    \label{fig:HistoricAdsTxtBusinessRelationshipsPopular}
	\end{minipage}
	\hfill
	\begin{minipage}[t]{0.32\textwidth}
    	\centering
    	\includegraphics[width=1\columnwidth]{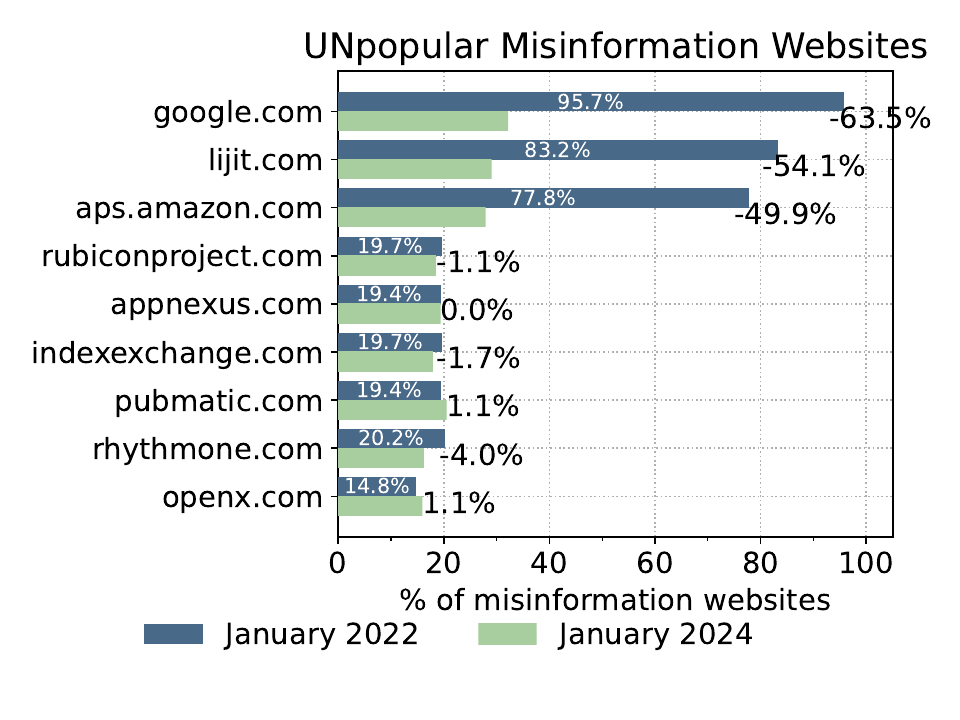}
    	\caption{Direct business relationships of top ad networks with \textbf{unpopular} misinformation websites after 2 years.}
    	\label{fig:HistoricAdsTxtBusinessRelationshipsUnpopular}
	\end{minipage}
        \hfill
	\begin{minipage}[t]{0.32\textwidth}
            \centering
            \includegraphics[width=1\columnwidth]{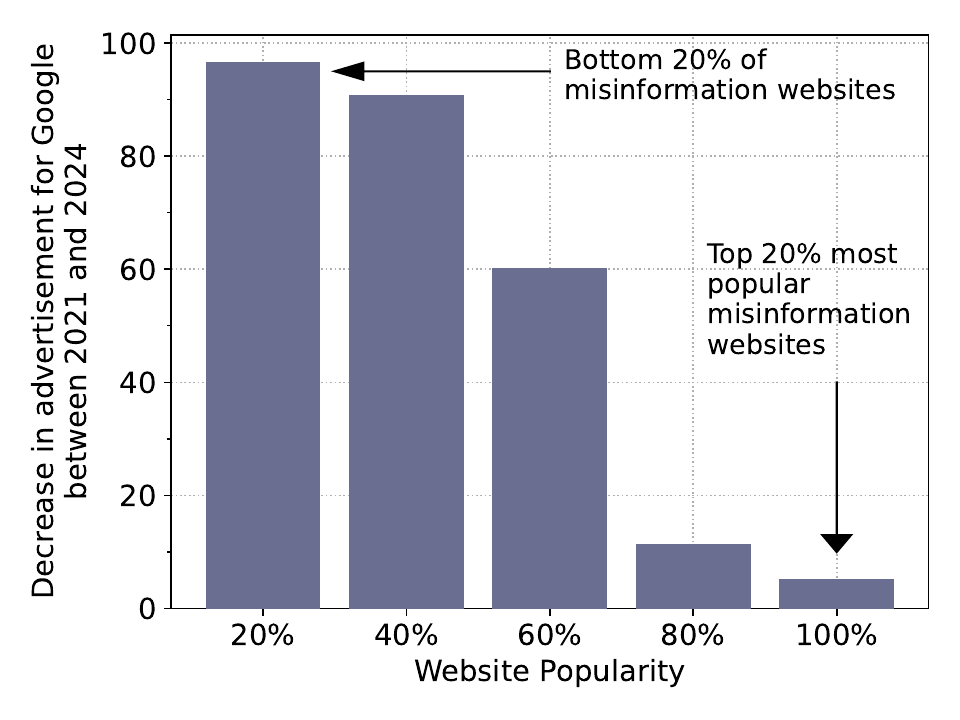}
            \caption{Difference in misinformation websites working with Google based on website network traffic.}
            \label{fig:popularityGroups}
        \end{minipage}
\end{figure*}

After we have built each persona, we use this user profile to visit the list of misinformation websites.
We utilize the open-source crawler presented in~\cite{papadogiannakis2022leveraging} and extend it to cover the needs of this work.
The crawler visits both the landing page and a randomly selected internal page of each website and detects the advertisers in these.
Prior work has demonstrated that internal pages can have significant differences in both structure and content (including ads and trackers)~\cite{10.1145/3419394.3423626}.
We extend the crawler to also detect online ads using cosmetic filters, based on Easylist~\footnote{\url{https://easylist.to/}} and uBlock Origin~\footnote{\url{https://ublockorigin.com/}}, two of the most popular ad-blocking projects.
Additionally, to make the crawler appear as close to an actual user as possible, we use a headful browser with custom preferences (\eg dark color scheme, window size, \etc).
Finally, we make use of Consent-O-Matic~\cite{nouwens2022consent}, a reliable extension, to automatically accept all cookies in consent banners and ensure that ads are not blocked because the user has not interacted with the consent banner.
We configure the extension to automatically accept all data processing purposes.

\section{Regulation Impact}
\label{sec:historicAnalysis}

\begin{figure*}[t]
    \begin{minipage}[t]{0.32\textwidth}
        \centering
        \includegraphics[width=1\columnwidth]{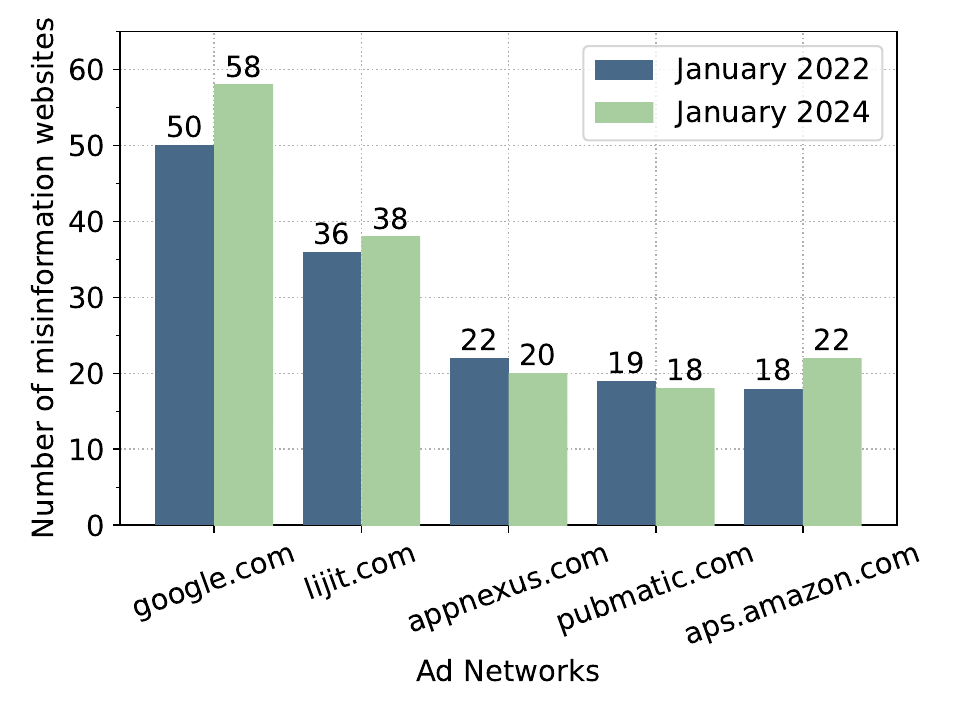}
        \caption{Evolution of popular misinformation websites explicitly stated in \sellersjson files of popular ad networks.}
        \label{fig:SellersJsonEvolution}
    \end{minipage}
    \hfill
    \begin{minipage}[t]{0.32\textwidth}
        \centering
        \includegraphics[width=1\columnwidth]{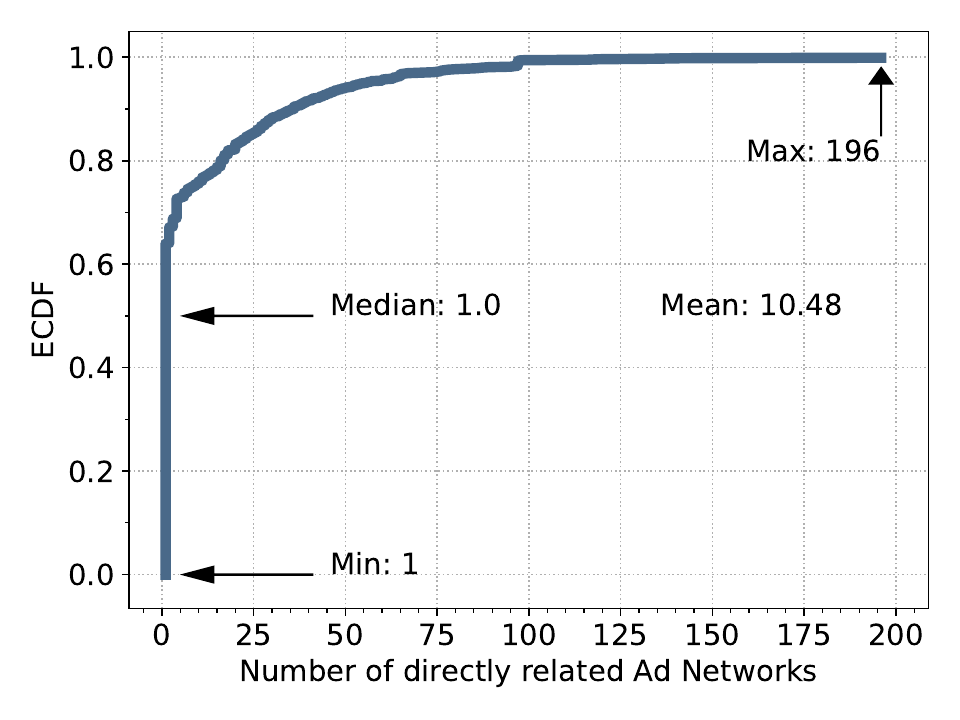}
        \caption{Distribution of the number of direct business relationships between a misinformation website and ad networks.}
        \label{fig:DirectRelationshipsDistribution}
    \end{minipage}
    \hfill
    \begin{minipage}[t]{0.32\textwidth}
        \centering
        \includegraphics[width=1\columnwidth]{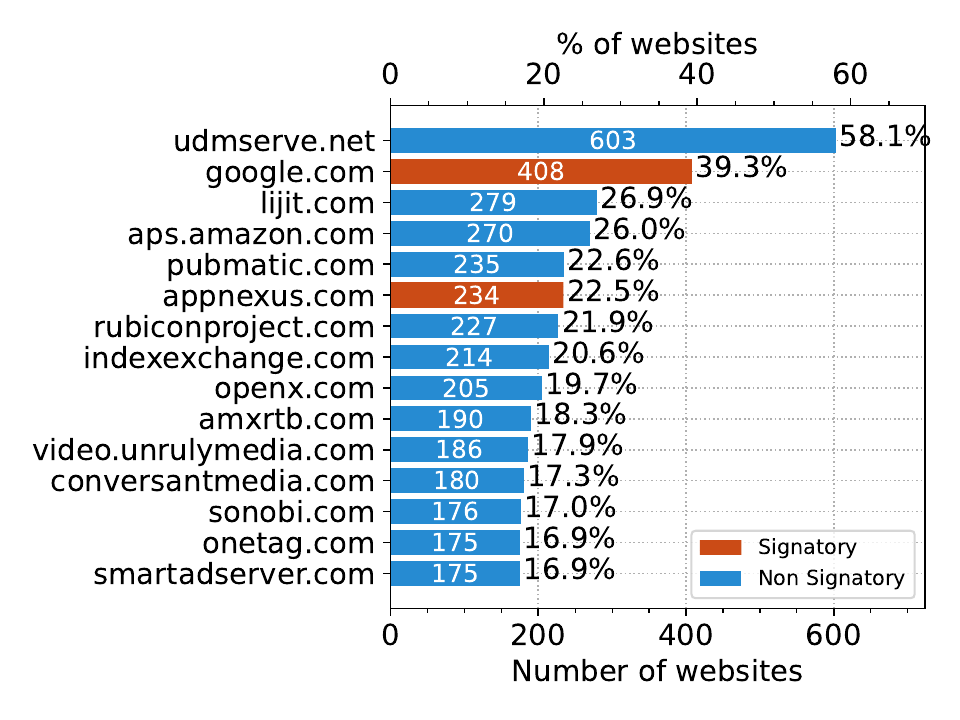}
        \caption{Direct business contracts with misinformation websites on July 2024 based on \adstxt and \sellersjson. Red bars represent \cop signatories~\cite{copodSignatories}.}
        \label{fig:AdsTxtSellersJsonBusinessRelationships}
    \end{minipage}
    \vspace{-0.2cm}
\end{figure*}

We set out to explore if the Strengthened Code of Practice on Disinformation has had a meaningful impact on the monetization of misinformation content.
We expect that the ecosystem will be improving because multiple signatories have committed to cutting the advertising revenue of disinformation spreaders~\cite{codeOfPractice}.
Inspired by prior work~\cite{papadogiannakis2023funds}, we study how direct business relationships of misinformation websites with ad networks have evolved over time.

We utilize the dataset of \adstxt and \sellersjson files described in Section~\ref{sec:dataset}, as well as the one collected in~\cite{papadogiannakis2023funds} for news websites in December 2021.
Arguably, publishers of fake news content could easily lie in their \adstxt files~\cite{papadogiannakis2023pooling,vekaria2023inventory}.
To discover the actual business relationships, we focus on information coming from both the publishers and the ad systems.
That is, we cross-reference the claims that publishers make in their \adstxt files with those of trustworthy ad networks in \sellersjson files.
We argue that since these are conflicting sides of the same coin, the intersection of their claims is the ground truth.
For each distinct DIRECT \adstxt entry, we review whether the respective publisher ID is also disclosed by the respective ad network.
For instance, if \emph{examplefakenews.com} has the entry \texttt{adnetwork.com, 12345, DIRECT} in its \adstxt file, we confirm that \emph{adnetwork.com} also discloses the identifier 12345 in its \sellersjson file.
Again, we focus only on \adstxt records labeled as \texttt{DIRECT} because they indicate a direct business contract between the misinformation publisher and the advertising system.
First, we study the behavior of the top ad networks and how their business relationships with misinformation websites have evolved over a period of 2 years.
We focus on the intersection of these datasets and find that there exist 461 misinformation websites with an \adstxt file in both datasets and with at least one business relationship also disclosed by the respective ad network.
We plot our findings in Figure~\ref{fig:HistoricAdsTxtBusinessRelationships}.

At first glance, most ad networks (as of January 2024) still work with misinformation websites, with only minor reductions.
6 out of the 9 most popular ad networks (\ie rubiconproject.com to openx.com in Figure~\ref{fig:HistoricAdsTxtBusinessRelationships}) have only blocked a very small percentage of misinformation websites (\ie up to 5.4\%).
However, some ad systems, including Google, Lijit, and Amazon, have stopped working with a significant portion of misinformation sites, dropping 51\%, 43\%, and 37\%, respectively, compared to two years ago.
While the drop in ad network relationships with misinformation websites aligns with the \cop, these websites represent a small fraction of total misinformation traffic.
To explore this, we use SimilarWeb~\cite{similarWeb} to gather traffic data for 95\% of the misinformation websites for which we have collected their \adstxt files, before and after the \cop.
In Figure~\ref{fig:MisinformationWebsitesNetworkTraffic}, we plot the distribution of their network traffic (number of visits on desktop and mobile) in January 2024.
The $x$-axis is in logarithmic scale, suggesting that there exist a lot of misinformation websites with low network traffic and a few highly popular that are able to attract millions of monthly visitors.
We split these websites on the 80\textsuperscript{th} percentile based on the observation that the top 20\% are popular websites, while the rest are unpopular.\footnote{The split 80-20 could have been made as 85-15, or 75-25. The point here is not to define an optimal threshold, but to create two categories: the ``popular'' misinformation websites and the ``unpopular'' ones.}
The 80\textsuperscript{th} percentile corresponds to a network traffic of approximately 290K monthly visits.
Consequently, we find 88 misinformation websites which we label as ``popular'', and 351 labeled as ``unpopular''.

We replicate the analysis from Figure~\ref{fig:HistoricAdsTxtBusinessRelationships} to track the evolution of ad network relationships with popular and unpopular misinformation sites in Figures~\ref{fig:HistoricAdsTxtBusinessRelationshipsPopular} and~\ref{fig:HistoricAdsTxtBusinessRelationshipsUnpopular}, respectively.
These figures paint a clear picture of the ad ecosystem.
Even though it seems that some ad networks no longer work with misinformation websites, they do so only for the unpopular ones.
Some of these websites have hardly any visitors and their effect on the dissemination of misinformation is limited, deeming them potentially insignificant.
However, for high-traffic misinformation sites attracting millions of visits, ad networks continue to partner with them, even a year and half after the \cop was signed (Figure~\ref{fig:HistoricAdsTxtBusinessRelationshipsPopular}).
Some top networks have severed ties with a small portion of the most popular misinformation websites, while others, like Amazon and Microsoft (AppNexus), now work with more popular misinformation websites than two years ago.

To further support our finding, we explore how business relationships with misinformation websites have been affected based on their popularity.
As a case study, we focus on Google, the most popular ad network~\cite{papadogiannakis2023pooling}.
We group misinformation websites into equal-sized buckets based on their monthly traffic in January 2024.
To validate these business relationships, we cross-reference them with Google's \sellersjson accounts, only reporting those that match.
We then compute the difference in misinformation websites working with Google between December 2021 and January 2024.
We express this difference as a percentage decrease $D = \frac{B-A}{|B|}\cdot 100$, where $A$ (\ie After) is the number of misinformation websites having a direct business relationship with Google on 2024/01, and $B$ (\ie Before) is for 2022/01.
Our findings, shown in Figure~\ref{fig:popularityGroups}, reveal a significant drop in verified accounts for less popular websites (left-most bar of the plot), with 96\% of websites ranked in the bottom 20\% based on traffic, no longer having a contract with Google.
On the other hand, we observe almost no difference for popular misinformation websites (\ie right-most bar of the plot).
Thus, we observe a discrepancy on how popular misinformation websites are treated, compared to unpopular ones.

Building on Figure~\ref{fig:popularityGroups}, we examine the correlation between a misinformation website's network traffic and whether an ad network stops working with it.
We make use of the traffic labels that MBFC assigns to news websites (``Minimal'', ``Low'', ``Medium'' or ``High'').
MBFC estimates traffic data through the SimilarWeb platform, but also takes into consideration the number of subscribers for print media and market size for TV/Radio entities~\cite{mbfcMethodology}.
We focus on the ad networks of Figure~\ref{fig:HistoricAdsTxtBusinessRelationships} that showed the greatest decrease in misinformation websites.
We find a correlation between the network traffic of a misinformation website and if it declares a business contract with an ad network after two years.
We present our findings in Table~\ref{tab:trafficCorrelation}, with a Spearman correlation coefficient of $\sim$0.6 and a $p$-value close to zero.
Over 70\% of misinformation websites with minimal network traffic no longer have a business relationship with these ad networks after two years.
We study how the popularity of misinformation websites has changed over time in Appendix~\ref{sec:popularity}.

\begin{table}[t]
    \centering
    \footnotesize
    \begin{tabular}{lrrrr}
	\toprule
    	\textbf{Ad Network} &
            \textbf{Websites} &
    	\textbf{Spearman} &
    	\textbf{p-value} &
    	\textbf{Minimal}\\
	\midrule
    	\emph{google.com} & 502 & 0.64 & 5.021135e-57 & -72.62\% \\
    	\emph{lijit.com} & 429 & 0.63 & 1.333227e-47 & -73.57\% \\
    	\emph{aps.amazon.com} & 415 & 0.64 & 8.267313e-48 & -75.28\% \\
	\bottomrule
    \end{tabular}
    \caption{Correlation between misinformation websites traffic and progress of direct ad networks business contracts. The last column shows the percentage of low-traffic websites that no longer have business relationships after 2 years.}
    \label{tab:trafficCorrelation}
    \vspace{-0.6cm}
\end{table}

Recent work has demonstrated that misinformation websites often use dark pooling in order to pool their ad inventory with that of lawful and acceptable websites~\cite{papadogiannakis2023pooling,vekaria2023inventory,vekaria2024turning}.
In an attempt to prove explicit relationships between popular misinformation websites and popular ad networks, we study their \sellersjson files.
In Figure~\ref{fig:SellersJsonEvolution}, we plot the number of misinformation websites from our list of $\sim$2,500 misinformation websites that are explicitly stated inside \sellersjson files of popular ad systems.
This is a definitive proof that ad networks still fund misinformation websites because they explicitly name them as verified ad inventory sellers.

We observe that the number of misinformation websites explicitly disclosed in the \sellersjson files of the most prestigious and popular ad networks is somewhat stable.
It is evident that the behavior of the top ad networks has not changed significantly over the last two years, even after the Code of Practice on Disinformation was signed.
In fact, studying the standard deviation of the number of popular misinformation websites in \sellersjson files (less than 2.5) shows that almost all networks have not stopped doing business with the misinformation websites.
We provide examples of such relationships in Appendix~\ref{popularFakeNewsExample}.

\begin{leftbar}
\noindent\textbf{Finding 1:}
Major ad networks still accept business relationships with misinformation websites, which account for the majority of the misinformation traffic.
After the \cop\/, these ad networks have ceased their collaboration mainly with the unpopular misinformation websites.
\end{leftbar}
\section{Financial Incentives}
\label{sec:adsTxtSellersJsonAnalysis}

The analysis of Section~\ref{sec:historicAnalysis} suggests that misinformation websites still rely on popular ad networks to monetize their content.
We cross-reference the accounts declared in \adstxt files with those found in \sellersjson ad networks, just like in Section~\ref{sec:historicAnalysis} and find that there exist 985 (87\%) misinformation websites with at least one substantiated business relationship with an ad network.
That is, for 985 misinformation websites, there exists an ad network that confirms the publisher account exists within its network and that it indeed belongs to a publisher.
This demonstrates that misinformation websites are still able to get approved by ad networks and monetize their content through ad revenue.

\begin{figure*}[t]
    \begin{minipage}[t]{0.49\textwidth}
        \centering
        \includegraphics[width=.8\columnwidth]{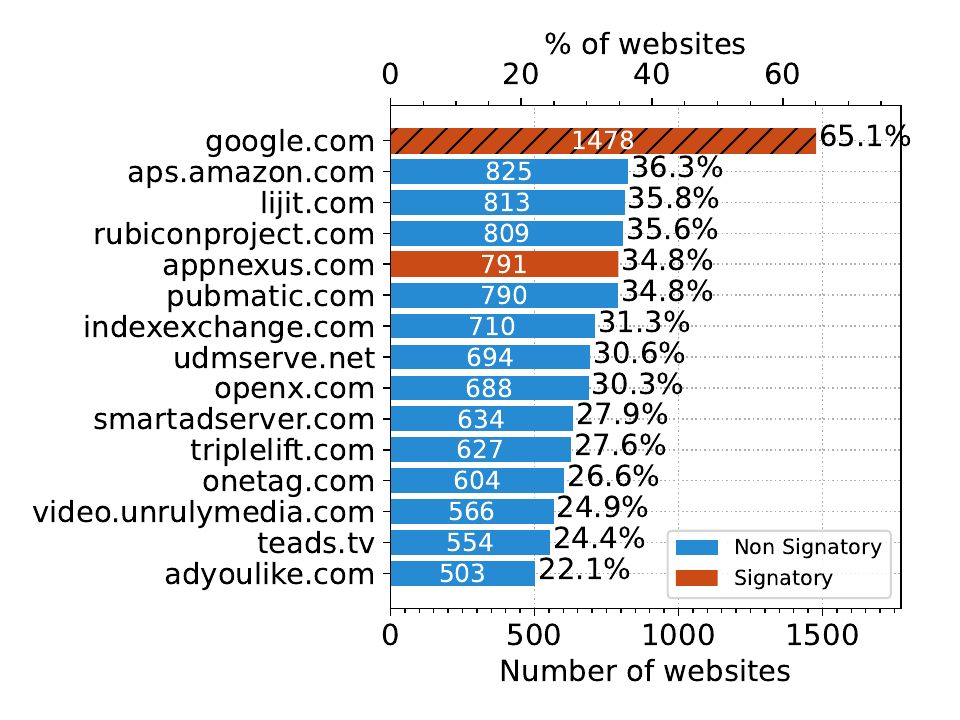}
        \caption{Direct business relationships with the most unreliable websites in July 2024 based on both \adstxt and \sellersjson files.}
        \label{fig:ScienceFeedbackAdsTxtSellersJsonBusinessRelationships}
    \end{minipage}
    \hfill
    \begin{minipage}[t]{0.49\textwidth}
        \centering
        \includegraphics[width=.8\columnwidth]{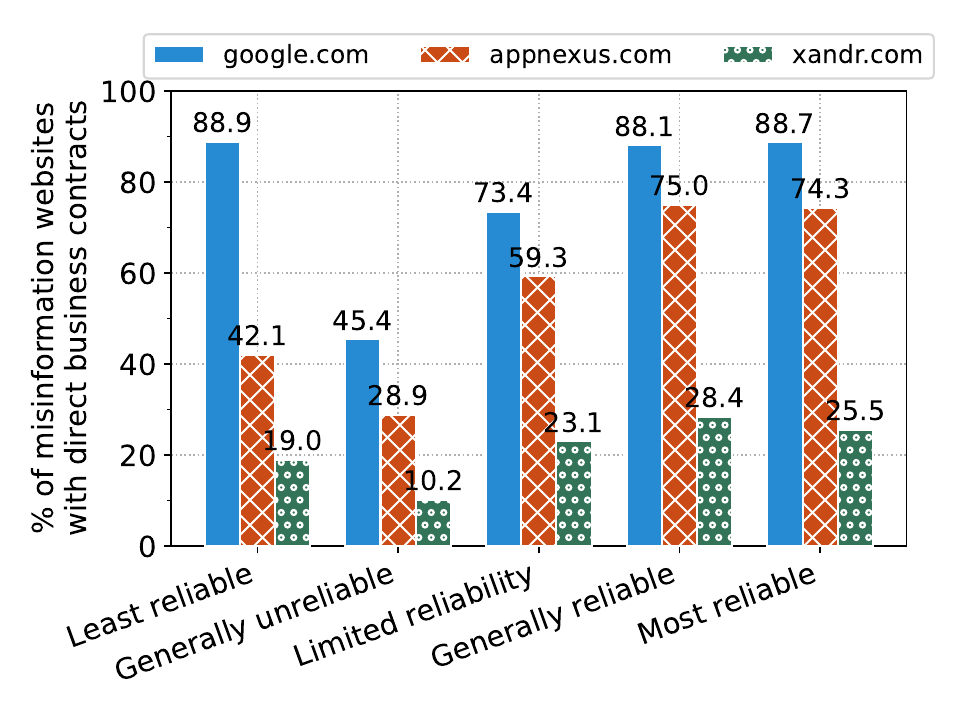}
        \caption{Business relationships among signatories of the \cop and news websites of various credibility ratings.}
    \label{fig:ScienceFeedbackAdsTxtReliability}
    \end{minipage}
    \vspace{-0.2cm}
\end{figure*}

\subsection{Direct Business Relationships}

We set out to explore what is the state of the advertising ecosystem two years after the Strengthened \cop was signed.
We collect the \adstxt files of misinformation websites and the \sellersjson files of ad networks on July 2024.
We plot in Figure~\ref{fig:DirectRelationshipsDistribution} the number of direct business relationships among misinformation websites and ad networks.
These relationships are extracted based on the ad accounts that both publishers and ad networks agree are valid through \adstxt and \sellersjson files, respectively.
We observe that the majority of misinformation websites have a direct contract with only a handful of ad networks.
However, there exist misinformation websites that form business contracts with tens or even hundreds of ad networks.
One such example is the website \emph{whatreallyhappened.com}.
This website serves an \adstxt file of over 8500 lines that contains verified \texttt{DIRECT} entries for 196 distinct ad networks.
In fact, the identifiers in this file are organized and labeled accordingly, indicating the presence of (dark) pools~\cite{papadogiannakis2023pooling,vekaria2023inventory}.

Next, we plot in Figure~\ref{fig:AdsTxtSellersJsonBusinessRelationships} the direct business relationships of top ad networks with misinformation websites.
We observe that for most ad networks, the behavior is similar; they have a DIRECT business relationship with about 16.9\%-58.1\% of misinformation websites whose \adstxt files we were able to obtain.
Even though these values might seem small, the underlying implications are significant.
They indicate that the most popular ad networks have a direct business contract with a large percentage of  misinformation websites, and that they facilitate the proliferation of misinformation content through ad revenue.
In fact, event the smallest value in Figure~\ref{fig:AdsTxtSellersJsonBusinessRelationships} suggests that ad networks have authorized at least 1 out of 5 misinformation websites that serve an \adstxt file.
The two ad systems that stand out are Google, and Underdog Media that controls \emph{udmserve.net}.
These ad networks directly partner with 400+ misinformation websites according to their \sellersjson files.
Even though their own guidelines prohibit monetizing false content~\cite{googlePublisherGuidelines,underdogMediaPublisherGuidelines}, they fail to filter out all such websites.
Altogether, there are hundreds of misinformation websites that still have direct business relationships with legitimate and popular ad networks and are able to generate revenue from dis-/mis-information content.
The accounts used to sell the ad inventory of these websites are recognized by popular ad networks.

\begin{leftbar}
\noindent\textbf{Finding 2:}
To this day, the most popular ad networks still have documented business relationships with over 500 distinct misinformation websites.
Ad networks that have signed the \cop have approved as ad inventory sellers more than 1 out of 3 misinformation websites that serve an \adstxt file.
\end{leftbar}

\subsection{News Websites Credibility}

To further support our finding that ad networks still work with misinformation websites to a large extent, we utilize the Science Feedback dataset on news websites credibility~\cite{scienceFeedbackDataset}.
The researchers behind this dataset have accumulated a list of over 20K news websites from more than 100 countries.
For each news website they assign a score reflecting its credibility, aggregating pre-existing and publicly available evaluations from trustworthy entities (\ie academic researchers, fact-checking organizations, expert raters).

We access the original list and successfully fetch the \adstxt files of $\sim$7K news websites in July 2024.
We perform a similar analysis as described earlier, but here, we make use of the list's original classification system and focus on websites labeled as ``Generally unreliable'' and ``Least reliable''.
We are able to extract 2,459 unreliable news websites with at least one \texttt{DIRECT} record in their \adstxt files and present our findings in Figure~\ref{fig:ScienceFeedbackAdsTxtSellersJsonBusinessRelationships}.
We discover a similar behavior with our previous findings.
Popular and legitimate ad networks have a direct business contract with hundreds of unreliable news websites, and this relationship is recognized by both the websites and ad networks.
Finally, in Figure~\ref{fig:ScienceFeedbackAdsTxtReliability} we focus on the signatories of the Code of Practice on Disinformation and explore their direct business relationships with news websites of various credibility scores.
We find that these ad networks work with all news websites and do not seem to have a distinctively different policy for unreliable news websites.
In the case of Google, we find that it works with almost as many of the least reliable websites (percentage wise) as with the most reliable.

\begin{leftbar}
\noindent\textbf{Finding 3:}
Two years after the Strengthened Code of Practice on Disinformation was signed, even the ad networks that signed the \cop do not considerably distinguish between unreliable and reliable news sources, allowing publishers to generate ad revenue from false claims.
\end{leftbar}
\section{Served Advertisements}

In previous sections, we demonstrate that the most popular ad networks still have business relationships with misinformation websites, and that the \cop did not have a substantial impact on the monetization of misinformation.
To validate our observations, we study ads that are actually served on misinformation websites.
Our goal is to explore to what extent ad networks still monetize misinformation content and if reputable advertisers might end up next to misinformation content.
This analysis follows the methodology described in Section~\ref{sec:ad-collection}.
Since misinformation websites are short-lived~\cite{chalkiadakis2021rise}, we focus only on active ones, detailed in Appendix~\ref{sec:activeWebsites}.

\subsection{Ad Networks}

Using data coming from 6 distinct user profiles (2 genders and 3 countries) and one clean user profile, we detect ads from 2,480 advertisers in 392 misinformation websites.
We attribute the large number of distinct advertisers to the use of different vantage points (\ie 3 geographic locations), thus being served ads from local brands.

First, we study the ad networks that deliver ads to misinformation websites.
We achieve this by studying the ad URLs the crawler detected.
We analyze these ad URLs and extract the domain from them, grouping them on the eTLD+1 level.
For example, ads served by \emph{adclick.g.doubleclick.net} and \emph{googleads.g.doubleclick.net} are all grouped under the \emph{doubleclick.net} domain.
To uncover the actual entities serving ads, we group domains based on the companies that own these domains.
For example, ads served by \emph{doubleclick.net} and \emph{googleadservices.com} are all attributed to Google LLC.
To identify entities that operate each domain, we utilize the DuckDuckGo Tracker Radar dataset~\cite{trackerRadar}.
We recognize that ad inventory (re)sales may involve intermediaries, but this is beyond the scope of this work.
We summarize our findings in Table~\ref{tab:adServingEntities}.

\begin{table}[t]
    \centering
    \footnotesize
    \begin{tabular}{lrr}
	\toprule
    	\textbf{Entity} &
            \textbf{\% misinformation} &
            \textbf{\# misinformation} \\
            & \textbf{websites} & \textbf{websites} \\
            \midrule
            Google LLC		            & 69.39\% & 272\\
            RevContent, LLC		        & 20.66\% & 81 \\
            AdRoll, Inc.		        & 12.50\% & 49 \\
            RTB House S.A.		        & 12.50\% & 49 \\
            Microsoft Corporation       &  8.93\% & 35 \\
            Criteo SA		            &  7.91\% & 31 \\
            MGID Inc		            &  7.14\% & 28 \\
            Taboola, Inc.		        &  6.12\% & 24 \\
            Zeta Global		            &  5.87\% & 23 \\
            Outbrain		            &  5.61\% & 22 \\
	\bottomrule
    \end{tabular}
    \caption{Ad network operators serving programmatic ads to misinformation websites. Even though Google and Microsoft have signed the \cop, they still deliver ads, and therefore revenue, to misinformation websites.}
\label{tab:adServingEntities}
\vspace{-.8cm}
\end{table}

We find that there are a lot of legitimate and popular ad networks that still deliver ads to misinformation websites.
Google is the most prominent ad facilitator of misinformation websites, serving ads to 69\% of the misinformation websites that contained ads.
The second most popular ad network that serves ads to misinformation websites is RevContent, known to be popular among misinformation websites~\cite{papadogiannakis2023funds,han2022infrastructure}.
Unfortunately, we find that ads are served on misinformation websites even by entities that have signed the Code of Practice on Disinformation (\eg Google and Microsoft)~\cite{copodSignatories}, suggesting that there is still work to be done.

It is important to mention that these findings highlight the issue described in previous sections with the analysis of \adstxt and \sellersjson files.
Ad networks may claim they once allowed misinformation websites (\ie findings of previous sections) but later blocked them.
Our analysis proves some still partner with misinformation publishers, serving ads and funding them.

To make matters worse, we discover ad networks serving misinformation ads.
For example, when our USA-based female persona visited the website \emph{theconservativebried.com}, Google served an ad for \emph{goop.com}, a website linked to pseudoscience and failed fact-check~\cite{mbfcGoop}.
We observe similar behavior with Outbrain, RevContent and Yandex LLC.
Evidently, ad networks do not properly review the content they promote, thus disseminating misinformation.
The \cop explicitly states that signatories (such as Google) should work towards stopping ads that contain false claims (Commitment 2: ``TACKLING ADVERTISING CONTAINING DISINFORMATION'').
All the above suggest that even though the \cop is a step towards the right direction, more effort is required to tackle the monetization and propagation of online misinformation.

\begin{leftbar}
\noindent\textbf{Finding 4:}
We discover that the most popular ad networks do not closely examine their ad placements.
They still serve ads, and thus ad revenue, to almost 400 misinformation websites.
Google and Microsoft, signatories of the \cop, place ads in 70\% and 9\% (resp.) of misinformation websites in our dataset.
\end{leftbar}

\subsection{Brand Safety of Advertisers}

Next, we analyze the actual advertisers that appear within the misinformation context.
We utilize the dataset of the previous section, but now focus on the landing pages users arrive when they click on ads displayed on misinformation websites.
We find that there exist a lot of advertisers whose ads are placed in dozens of different misinformation websites.
This behavior is consistent for various types of ads and for all personas.
For example, using the USA-based female persona, we find that \emph{jjshouse.com}, a popular online store for wedding dresses, was advertised in 80 distinct misinformation websites.
Similarly, the Greece-based male persona was served ads from Temu in 86 distinct misinformation websites (22\%) while ads from \emph{overnightglasses.com} were served to the male US-based user in 59 distinct misinformation websites.
We should note that this finding is not a finger pointing exercise for these advertisers and does not indicate that these advertisers choose to appear in misinformation websites.
Most likely, these ads are placed next to misinformation because of the ad network believing they are of interest to the virtual user.
However, we should note that this can have a big impact on advertisers (\ie brand safety) if visitors associate misinformation content with specific services or products.

Finally, we analyze the significance of the advertisers found in misinformation websites. 
To our surprise, we find ads from very reputable companies appearing in misinformation websites.
We provide screenshots of examples~\cite{openSourceData}.
We find 23 Fortune 500 companies advertising in misinformation websites, threatening their reputation and consumer trust due to negative brand associations.

\begin{leftbar}
\noindent\textbf{Finding 5:}
Ads from very popular and renowned brands (even from 23 of the largest companies in the world) are served in misinformation websites.
Users can see an ad of a specific brand in up to 86 distinct misinformation websites.
\end{leftbar}
\section{Related Work}

The monetization of misinformation content through online ads has been thoroughly studied.
In fact, in~\cite{Thirumuruganathan2021}, authors performed a cost-based analysis on fake news interventions and proposed a model that can help combat the proliferation of misinformation websites.
In~\cite{bozarth2021market}, authors studied fake news and low-quality news websites and found that fake news publishers monetize their content using reliable ad servers.
Similarly, in~\cite{Broniatowski2023}, authors studied the monetization schemes of misinformation websites, focusing on anti-vaccine pages, demonstrating they mainly rely on ads.
In~\cite{chalkiadakis2021rise}, authors found that some popular third-party services have higher presence in fake news sites compared to real news.
In~\cite{han2022infrastructure}, expanded upon this finding and found that misinformation websites rely on popular Web services such as Cloudflare, Google and RevContent to host and monetize their content. 
This behavior is leveraged by~\cite{papadopoulos2023fndaas}, where authors proposed a content-agnostic fake news detection service using network and structural characteristics (including requests towards third-party advertising services), and~\cite{hounsel2020identifying}, authors utilized non-perceptual features (\eg domain name, DNS config) to train a multi-class model that detects disinformation websites.

Regarding the \adstxt and \sellersjson standards that are the foundation of our methodology, in~\cite{bashir2019longitudinal}, authors performed a longitudinal study of \adstxt files and their adoption.
In~\cite{vekaria2023inventory}, authors also measured the compliance of the ad ecosystem with ad standards and explored how publishers use \adstxt files to form dark pools, a way to circumvent restrictions and monetize misinformation content.
Similarly, in~\cite{papadogiannakis2023pooling}, authors further explored dark pooling along with hidden intermediaries and highlighted the importance of limiting the ad revenue generated by misinformation publishers.
In~\cite{vekaria2024turning}, authors demonstrated that dark pooling can be diminished by notifying stakeholders and ad networks.
In this work, we expand the methodology and make use of the dataset presented in~\cite{papadogiannakis2023funds}, where authors processed \adstxt and \sellersjson files to study the business relationships between fake news websites and ad networks.
They discovered that fake news websites rely on popular and credible ad networks to generate revenue.

Previous work has examined the content of ads served on misinformation websites, finding that nearly half of the ads were problematic~\cite{zeng2020bad} and that some advertisers tolerate fake news websites due to lower costs~\cite{braun2019fake}.
In this work, we mainly focus on the ad networks responsible for serving ads on misinformation websites and only briefly examine the advertisers to discuss brand safety.
Brand safety has been more extensively studied in~\cite{ahmad2024companies}, where authors demonstrated that ads largely finance misinformation websites and that can have a great effect on advertisers' brand safety.
Previous work has also studied how questionable content affects the brand of pre-roll advertisers~\cite{bellman2018brand} and advertisers that utilize programmatic advertising~\cite{shehu2021risk}, and how reputable brands can appear next to illicit content, unbeknownst to them~\cite{papadogiannakis2024ad}.
\section{Discussion \& Conclusion}

\subsection{Discussion}

The Strengthened Code of Practice on Disinformation is a step towards the right direction, but more work is needed.
This study doesn't aim to discredit the \cop but to stress the need for better scrutiny of ad placements and more misinformation websites marked ineligible for monetization.
We believe ad networks must ensure misinformation is no longer profitable, preventing even popular websites from surviving on ad revenue.
Reducing financial incentives for misinformation can improve information quality and trust in factual reporting.
Although the self-regulatory nature of the \cop is promising, its adoption is limited, with only 44 signatories as of July 2024.
It is evident that current signatories have been laying the foundation, but we hope new signatories from the online ad ecosystem (\ie ad networks) will take initiative for more substantial actions.

It is customary in such studies to have a control group, that is, a set of non-disinformation websites and study what is the change in advertisements in these websites as well.
Such a control group enables researchers to argue whether a change they see in misinformation websites is due to the \cop.
However, this is not needed for this work.
Indeed, we are not interested in attributing any changes we see to the \cop (or not).
The goal of this work is to discover if the change that the \cop expected (\ie that disinformation websites will not have ads) happened and to what extent.

Finally, we acknowledge the argument that some ad networks might still work with misinformation websites because they classify these websites as serving ``low credibility content'' rather than deliberate disinformation.
Classifying a website as misinformation or not falls outside the scope of this work.
We rely on external sources (\eg MBFC, ScienceFeedback) that work on this exact topic and the academic community has already reviewed their reliability.

\subsection{Conclusion}

The Strengthened Code of Practice on Disinformation was signed in June 2022, in an attempt to fight disinformation in a self-regulatory manner.
The first commitment of the \cop is that entities involved in ad placements demonetize dissemination of disinformation.
The expectation is that since the \cop was signed by leading industry players (\eg Google and Microsoft), two years after the introduction of the strengthened \cop, things will have improved.
In this work, we perform a historical analysis and demonstrate that this is not the case.
In fact, misinformation websites are still able to retain business contracts with popular ad networks and monetize their content.
We show that the adoption of the Code of Practice for Disinformation has not made a significant impact on the misinformation ecosystem, and that ad networks stopped working with only insignificant misinformation websites that do not have a lot of visitors. 
Finally, we perform a dynamic analysis of ads served on misinformation websites.
We establish that not only popular ad networks still facilitate the monetization of the majority of misinformation websites, but that they also serve ads of respectable and popular brands next to misinformation content.
\begin{acks}
Funded by the European Union.
Views and opinions expressed are however those of the author(s) only and do not necessarily reflect those of the European Union or European Research Executive Agency (REA). Neither the European Union nor the granting authority can be held responsible for them.
\end{acks}

\bibliographystyle{ACM-Reference-Format}
\balance
\bibliography{main}

\appendix
\section{User Profiles}
\label{sec:personas}

We form browser profiles to emulate real and distinct Web users.
Our methodology revolves around the idea that user profiles can be formed based on websites that are representative of the user's gender~\cite{agarwal2020stop}. 
For example, if websites \emph{a.com}, \emph{b.com} and \emph{c.com} are mainly visited by female users from the Netherlands, and we visit these websites with a Dutch IP address using a clean user profile, third parties will deduce that the profile belongs to a female user from the Netherlands.

To form a list of websites that are associated with a gender of a specific country, we utilize the SimilarWeb analytics platform~\cite{similarWeb}.
Specifically, we visit SimilarWeb and extract the 5 most popular websites in each category and each country of our experiment.
SimilarWeb uses a taxonomy system with 210 distinct website categories (including subcategories).
For each website, we extract its demographic data (if available).
SimilarWeb uses two gender groups (\ie female and male) to classify visitors of websites.
Next, we sort websites based on their audience’s gender representation and ensure that all websites have a representation of at least 55\% for that gender.
We set an upper threshold of 200 websites and a lower threshold of 100 websites to train each persona.
The lower threshold is more than enough, since previous work has demonstrated that a persona is stable after visiting the top 10 websites associated with users of that demographic~\cite{agarwal2020stop}.
This process leads to 6 (combination of 2 genders and 3 countries) lists of websites that represent the interests of the specific persona.

We utilize Playwright~\cite{playwright}, a browser automation framework and set up legitimate browser profiles, retaining cookies and local storage.
To build the user profiles, we try to emulate the behavior of a real user that visits some websites to consume online content.
We try to make the behavior of our automated crawler as non-deterministic as possible to appear as a legitimate user.
The behavior we are emulating is that of a user that visits a website, reads its content, and then navigates through the website after quickly inspecting the content of each page.
For each website in our target list, we
(i) visit the landing page,
(ii) wait for a random number of seconds in range [2, 5], and
(iii) explore 5 subpages.
In order to explore a subpage, our crawler clicks on a randomly selected internal URL and navigates to this subpage, waits for a random number of seconds to emulate a user consuming content, scrolls for a random number of seconds and then waits for an additional random number of seconds.
The websites of the list are visited in a random order and after all websites have been visited, the user profile is stored for further use.
\section{Popularity Evolution}
\label{sec:popularity}

We find that popular ad networks no longer work with websites with low traffic.
The classification of a website's popularity is based on traffic data from 2024.
We study if their popularity was similar when the ad networks made their decisions.
We compare the popularity rank of the 461 misinformation websites of Figure~\ref{fig:HistoricAdsTxtBusinessRelationships}.
We make use of the Tranco list and compare their ranks during June 2021 and June 2023 (\ie one year before and one year after the \cop was signed).
We find that for the websites in both lists, their rank distribution is somewhat similar.
In fact, the mean rank decreased by -9\% (\ie more popular) during this 2-year period, while the median rank increased 42\%.

Next, we perform a 2-sample KS test for the ranks of misinformation websites in June 2021 and June 2023 and find that the p-value is large enough (\ie 0.16) so we cannot reject the hypothesis that the two distributions are the same.
In addition to this, we argue that the difference in the median rank might be credited to some extent the variance the Tranco list exhibits from one version to the other.
We perform exactly the same experiment for the same time period for the real news websites published in~\cite{papadogiannakis2023funds}.
We find that for real news websites, the median rank increased by 35\%, a comparable magnitude.
Altogether, we demonstrate that the popularity of the misinformation websites we studied has not changed significantly before and after the \cop was signed, thus our argument that ad networks have dropped only unpopular misinformation websites still holds.
We selectively compare the rank of misinformation websites in June 2023 and not January 2024 (just like we do in Figure~\ref{fig:HistoricAdsTxtBusinessRelationships}) because Tranco has since changed the ground truth datasets they utilize to compute the aggregated ranks.

\section{Popular Misinformation Websites}
\label{popularFakeNewsExample}

To our surprise, we find that even today, there are a lot of ad networks working with certain misinformation websites that one can hardly argue they are legitimate.
For example, we find that there are at least 10 of the most popular ad networks (including Google, Amazon, AppNexus and Pubmatic) that have an explicit and direct business relationship with \emph{worldlifestyle.com}.
According to MediaBias/FactCheck, WordLifestyle is ``a strong pseudoscience website based on the promotion of false or misleading claims regarding science and health'' and it has a low factual reporting score ``due to a lack of sourcing and blatant clickbait abuse to generate advertising revenue''~\cite{mbfcWorldLifeStyle}.
Even such websites that are known to take advantage of the ad ecosystem in order to monetize misinformation still have direct business contracts with popular ad networks.
In fact, by manually visiting this website, we find that Google indeed serves ads to visitors.

In addition to this, we find that the website of ``One America News Network'' (OANN) is an approved ad inventory seller within the most popular ad networks (including Google and AppNexus, signatories of the Strengthened Code of Practice on Disinformation).
OANN is considered a far-right biased news website that promotes propaganda, conspiracy theories and has over 10 failed fact checks~\cite{mbfcONNN}.
According to SimilarWeb, this website achieves over 2.5M monthly visits, and in conjunction with the fact that almost 95\% of the visitors come from the United States, an advertising valuable country, it can generate substantial ad revenue.
\section{Active Misinformation Websites}
\label{sec:activeWebsites}

Inspired by previous findings that misinformation websites might have a short longevity~\cite{chalkiadakis2021rise}, we aim to refine our list by accumulating only misinformation websites which are still active.
Towards this extent, we perform a manual analysis using independent reviewers.
We start from the extended list of 2,469 misinformation websites and using an automated crawler we take screenshots of the entire landing page of each website.
Our crawler was able to successfully access 2,049 websites and capture screenshots.
Next, two independent reviewers coming from a computer science background (not the authors of this work) evaluate which websites are still active based on the content of the landing pages.
Reviewers are allowed to classify each website as either ``Active'' or ``Inactive''.
Moreover, reviewers are allowed to manually visit a website and navigate through its pages if they are not certain which label to assign simply based on the screenshot of the landing page.
The two reviewers are independent of each other and completed the task in a different environment, without influencing one another.

While manually rating websites, the reviewers noticed that some domains were active in the sense of responding to requests, but no longer served news content.
This was mainly because these websites were up for sale (\ie parked domains) or because they served empty content.
In both of these cases, reviewers classified such websites as inactive.
Domains that did not respond to HTTP(S) requests, domains that were non-existent (\ie \texttt{DNS NXDOMAIN}) and domains that timed-out when accessed were all labeled as inactive.

Both reviewers evaluated the entire list of 2,469 misinformation websites and agreed to include 1,889 websites, agreed to exclude 568 websites and disagreed on 11 websites.
The reviewers agreed in 99.55\% of websites and had a Cohen Kappa score of 0.99.
Note that a high inter-rater agreement score is expected in this context because deciding if a website is active is an easy task for a human evaluator.
We decide to only accept websites that both raters agree are active and end up with a list of 1,889 active misinformation websites.
We make this list of active misinformation websites publicly available~\cite{openSourceData}.
\section{Ethical Considerations}
\label{sec:ethics}

In this work, we make concentrated effort to not affect the performance of any Web service in any way.
To download \adstxt and \sellersjson files, we utilize existing crawlers and reach each domain only once to download the specific file.
This behavior is far less intrusive compared to the automated processes that DSPs and SSPs use daily to verify programmatic advertising.
Additionally, our data was collected once and all of our analysis is performed offline.
Regarding the collection of ads, we ensure that we click on each ad URL only once, similarly to methodologies presented in previous work~\cite{papadogiannakis2023funds,vekaria2024turning}.
Since previous work has demonstrated that the cost per thousand ad impressions is only a few dollars~\cite{pachilakis2021youradvalue}, we argue that our study's influence on the advertising ecosystem is negligible.
Finally, following the GDPR and ePrivacy regulations, we do not collect or process any personal information of real users.

\end{document}